\documentclass[12pt]{iopart}

\usepackage[utf8]{inputenc}
\usepackage[T1]{fontenc}
\usepackage{epsfig,xcolor,graphicx,multirow, physics,soul}
\usepackage[normalem]{ulem}

\def\T{\textit{\textbf{T}}}

\begin{document}

\title[Flow of asymmetric elongated particles]{Flow of asymmetric elongated particles}

\author{Viktor Nagy,\textit{$^{1}$}
Bo Fan,\textit{$^{1,2}$}
Ell\'ak Somfai,\textit{$^{1}$}
Ralf Stannarius,\textit{$^{3}$}
and
Tam\'as B\"orzs\"onyi\textit{$^{1}$}
} 

\address{
$^1$Institute for Solid State Physics and Optics, Wigner Research Centre for
Physics, P.O. Box 49, H-1525 Budapest, Hungary\\
$^2$Physical Chemistry and Soft Matter, Wageningen University \& Research, Wageningen, The Netherlands \\
$^3$Institute of Physics, Otto von Guericke University,  Universit\"atsplatz 2, 39106 Magdeburg, Germany}

\ead{borzsonyi.tamas@wigner.hu}
\vspace{10pt}

\begin{abstract}
Shear induced orientational ordering of asymmetric elongated particles is investigated experimentally. Corn grains and pegs with one end sharpened are studied using X-ray Computed Tomography (CT) during quasistatic shearing and silo discharge. 
We show that asymmetries can be detected in the orientational distributions of the particles, which are related to the modulated rotation of the particles during shear flow. 
Namely, when the particles rotate in a plane that is not horizontal, they spend more time with the sharper (lighter) end pointing up, which can be explained using energetic arguments. We quantify the resulting asymmetry of the orientational distribution in a split bottom Couette cell and in a silo discharge process.
\end{abstract}

\section{Introduction}\label{sec1}

When a granular material consisting of elongated particles is subjected to shear, the particles get orientationally ordered. The average alignment of the particles' long axis is nearly parallel to the flow, it is characterized by a small angle $\theta_\text{av}$, which decreases with increasing particle elongation as it was shown in numerical simulations \cite{ankireddyPRE2009,campbellPOF2011,guoJFM2012,nadlerPRL2018,borzsonyiPRL2012,borzsonyiPRE2012} 
and laboratory experiments \cite{borzsonyiPRL2012,borzsonyiPRE2012,borzsonyiSM2013,wegnerSM2012,guillardSR2017}.
Several works have shown that for elongated or flat particles increasing shape 
anisotropy of the grains leads to stronger orientational ordering, higher effective 
friction of the sample or stronger shear banding 
\cite{borzsonyiPRL2012,marschallGM2020,botonPRE2013,botonEPJE2014,nagyNJP2020,hidalgoPRF2018,trulssonJFM2018,nouguierCRM2010}.
For particles with low interparticle friction ($\mu_p\leq0.4$ in 3D and $\mu_p\leq0.15$ 
in 2D) an interesting non-monotonic dependence of the effective friction was found as 
a function of grain  elongation \cite{nagyNJP2020, trulssonJFM2018}.
The effect of grain shape on the flow field in a silo, the discharge rate and the clogging 
probability was also investigated
\cite{szaboPRE2018,taoCES2010,markauskasGM2011,clearyCSIRO1999,clearyAMM2002,liuPT2014,liCES2004,langstonCES2004,ashourSM2017,escuderoEPJWC2021,calderonPT2017,reddyJSM2021}.
For elongated grains, the flow field was found to be more concentrated to the silo center 
and had larger temporal fluctuations compared to the case of spherical grains
\cite{szaboPRE2018,taoCES2010,markauskasGM2011}.
Even if some of the above investigations involved asymmetric elongated grains e.g. corn seeds, sesame seeds \cite{taoCES2010,gonzalezCES2011,escuderoEPJWC2021,calderonPT2017} or dumbbells \cite{reddyJSM2021}, the role of grain shape asymmetry is not well explored in shear flow or silo flow.

In the present work we use two types of elongated asymmetric grains: sharpened pegs and hard (dry) corn seeds in a shear flow and in silo flow. We use a non-invasive method (X-ray computed tomography) 
to detect the location and orientation of each particle inside our 3-dimensional experimental system. We analyze the orientation distributions and focus on effects related to the asymmetric shape of the grains.

\section{Experimental system}
\label{exp}

\begin{figure}[!htp]
\begin{center}
  \includegraphics[width=\columnwidth]{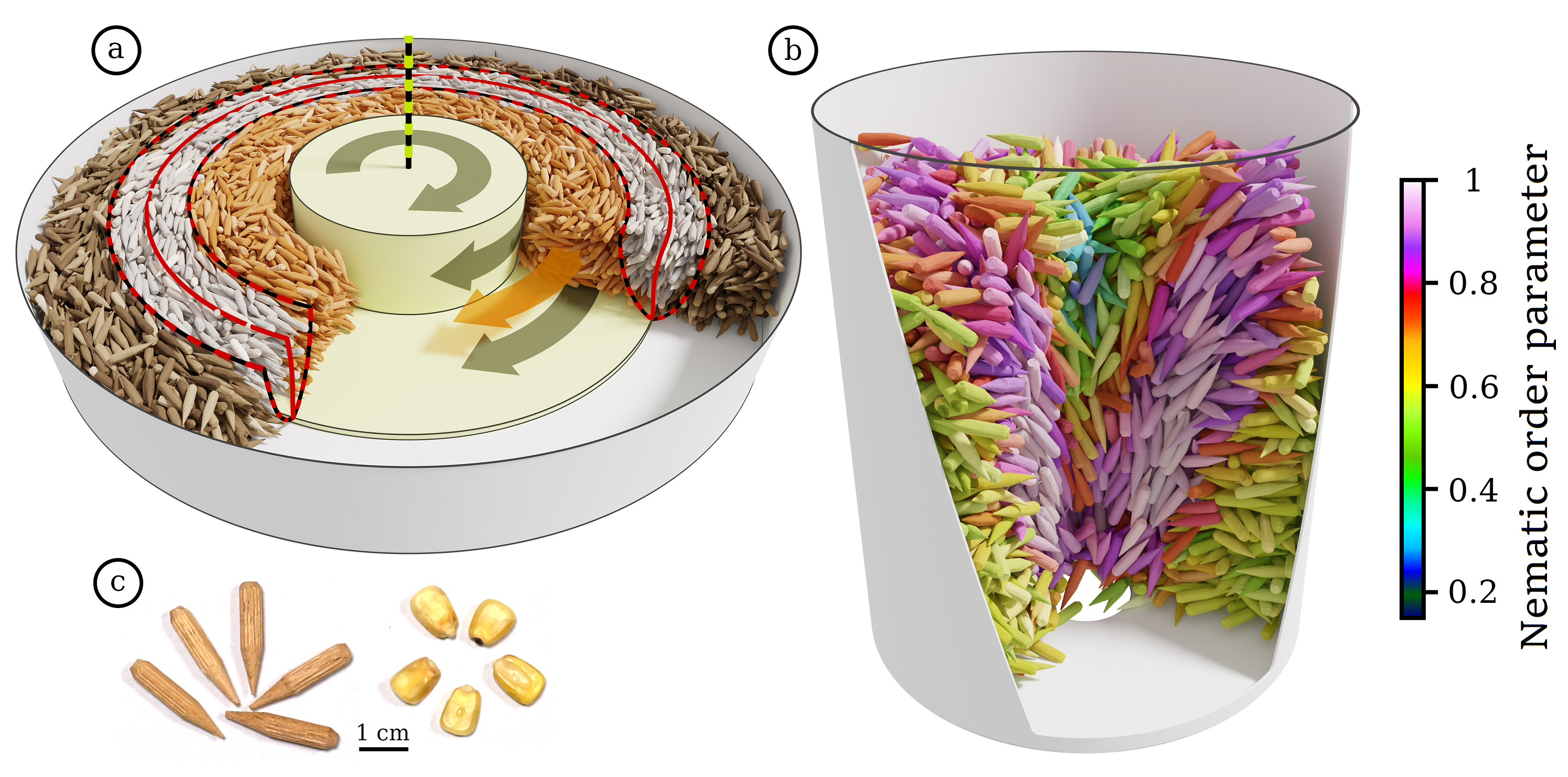}
\end{center}
  \caption{Sketches of the experimental configurations: (a) cylindrical split-bottom
shear cell (b) silo. Both pictures include grains as detected in tomograms.  
In panel (a) the shear zone is indicated with red lines, while particles in the shear 
zone are colored white.
In panel (b) particle colors represent the local nematic order parameter, according to the 
colorbar next to the cylinder. 
(c) Photographs of the samples: sharpened wooden pegs and corn seeds.
  }
  \label{setup}
\end{figure}

In this work, two experimental geometries have been employed to study the flow of a granular
assembly of elongated asymmetric particles. In the first experiment, the granular material was 
exposed to shear in the so called cylindrical split-bottom shear cell (see Fig. \ref{setup}(a)). 
In this device, the central part of the granular sample is rotated, while the outer part of the 
sample is not moving. The sheared region is between the moving and standing parts, with 
the highest shear rate in the middle of this region. We define the core of the shear zone as 
the region where the time averaged rotation rate is in between $5 \%$ and $95 \%$ of the 
rotation rate of the inner part. 
This shear zone core is indicated with red contours and 
white particles in  Figs. \ref{setup}(a), \ref{orientations}(a) and a dashed red line in Figs. \ref{shear-rod-corn}(a-f). The width of the shear zone is expected to scale linearly with particle size and to be larger for spherical particles than for irregular ones \cite{fenisteinPRL2004}. As we will see later, the geometry of the shear zone in our experiments was very similar for the two types of grains we used.
The radius of the cell was 28.5 cm, the radius of the rotating plate was 19.5 cm, while the 
height of the granular layer was 6 cm. Before recording the data, the sample was 
pre\-sheared with 15 full rotations, in order to eliminate any transient effects and to ensure 
stationary shear throughout the experiment. During the experiment, 
quasistatic shearing was applied which was stopped at regular intervals of $360^\circ$ 
of rotation of the inner part, and then a pair of X-ray computed tomograms (CT) was taken of the sample.
The plate was rotated by a small amount ($2^\circ$) between the two tomograms, with the aim 
of enabling the detection of particle displacements for the determination of the flow field.  
We recorded 50 pairs of tomograms for sharpened wooden pegs and 19 pairs of tomograms for corn seeds.
In the second experiment, the granular material was discharged from a nearly
cylindrical silo with a circular orifice at the bottom (Fig. \ref{setup}(b)). 
The diameter of the orifice was relatively
small so that the flow often clogged. Each clogged configuration was recorded with X-ray CT.
We recorded 109 tomograms for sharpened pegs and 30 tomograms for corn seeds.
The container had a height of 21.4 cm and a diameter of 19 cm. The tomograms for both experiments 
were obtained with the Siemens Artis zeego X-ray Tomograph of the STIMULATE-lab of the Otto von 
Guericke University in Magdeburg. The recorded volume was 25.2 cm $\times$ 25.2 cm $\times$ 19 cm, 
with a spatial resolution of 2.03 pixel/mm, resulting in tomograms of 512 $\times$ 512 $\times$ 
386 pixels. The recording of a single tomogram took about 2 minutes.

The particles used here share a common feature: in addition to being elongated, both are asymmetric 
in the sense that the two ends are different: the corn seeds have a wedge shape, while for the pegs 
one of the ends is sharpened (see the photographs in Fig. \ref{setup}(c)). The pegs had a diameter 
of $d = 5$ mm and a length of $L = 25$ mm (aspect ratio: L/d=5), while the typical dimensions of the corn seeds were about $d_1 = 5$ mm, $d_2 = 8$ mm,  and $L = 12$ mm (aspect ratio: $2L/(d_1+d_2)=1.8)$.
The particle positions and orientations were obtained by tailored 3D image processing.  After an initial adaptive binarization of the 3D absorbance field obtained by the X-ray CT measurements, particles were separated by binary erosion and reconstructed by subsequent regrowth.
The pictures in Fig. \ref{setup}(a,b) show reconstructed pegs from actual measurements.
In the shear cell (Fig. \ref{setup}(a)) the particles are colored according to their
location in order to visualize the core of the shear zone. In the silo (Fig. \ref{setup}(b)), 
the colors represent the local nematic orientational order parameter, which was obtained by averaging over particles in rings of similar height and radius from the central axis.

\section{Results and Discussion}
\subsection{Shear flow}

In the first experiment, we characterize the flow field and the orientation of the particles in 
stationary shear flow in the split bottom shear cell. 
\begin{figure}[!ht]
\begin{center}
  \includegraphics[width=\columnwidth]{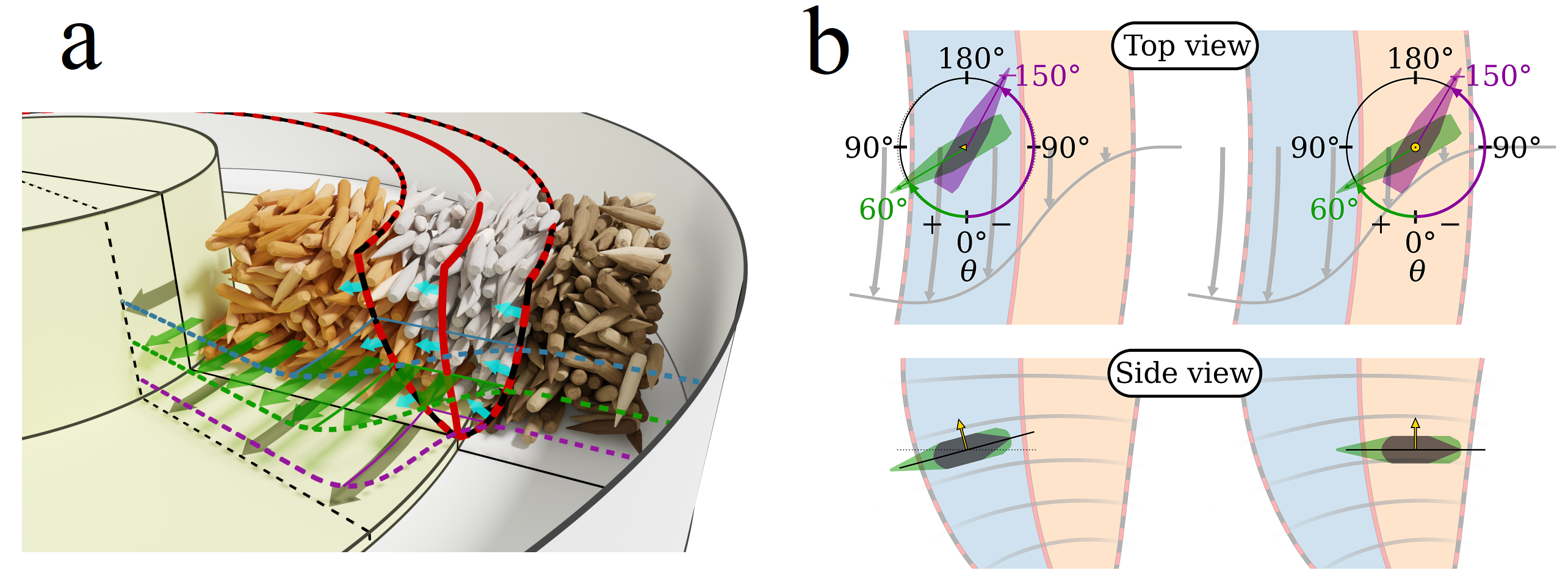}
\end{center}
  \caption{Geometry of the shear flow in a cylindrical split bottom cell. In panel (a) the
contours and the middle of the shear zone core are indicated with red lines, the particles in the 
shear zone core are colored white. The velocity profiles at three different heights are indicated with
three different colors, the velocity gradient at the same heights is shown with turquoise arrows, and the green arrows indicate the velocity profile in the middle of the granular layer. 
In panel (b) a top view and a side view of the shear zone is sketched. In the top view the flow 
velocity is indicated by arrows. In the side view, the direction of the local velocity gradient 
is indicated by lines: it is nearly horizontal in the outer half of the shear zone (orange) 
and slightly tilted in the inner part of the shear zone (light blue).
Two particles lying in the shear plane with orientation angles $\theta=-150^{\circ}$ and 
$\theta=60^{\circ}$ are shown as examples.
  }
  \label{orientations}
\end{figure}
Figure~\ref{orientations} indicates the main features of the 
velocity field in this system: the flow velocity is strictly tangential, while the velocity gradient is 
nearly horizontal. 
The side view in Fig. \ref{orientations}(b) shows that the lines parallel 
to the local velocity gradient are somewhat curved. This results in a practically horizontal 
velocity gradient in the outer half of the shear zone core (filled with orange) and a slightly
tilted velocity gradient in the inner half of the shear zone core (filled with light blue).
As discussed in earlier works \cite{borzsonyiPRL2012,borzsonyiPRE2012,borzsonyiSM2013,wegnerSM2012},
elongated particles in a shear flow get oriented with their 
long axis predominantly in the shear plane (spanned by the flow velocity and 
velocity gradient). They perform a rotation within the shear plane with a fluctuating rotation
velocity which depends on the actual interactions with neighbours. The ensemble averaged rotation 
velocity, however, shows a clear orientation dependence: particles with their long axis nearly 
parallel to the flow rotate slower than those perpendicular to the flow. This results in 
orientational ordering with an average alignment of the longest axis of the grains nearly parallel to the main flow.

\begin{figure}[!htp]
\begin{center}
  \includegraphics[width=\columnwidth]{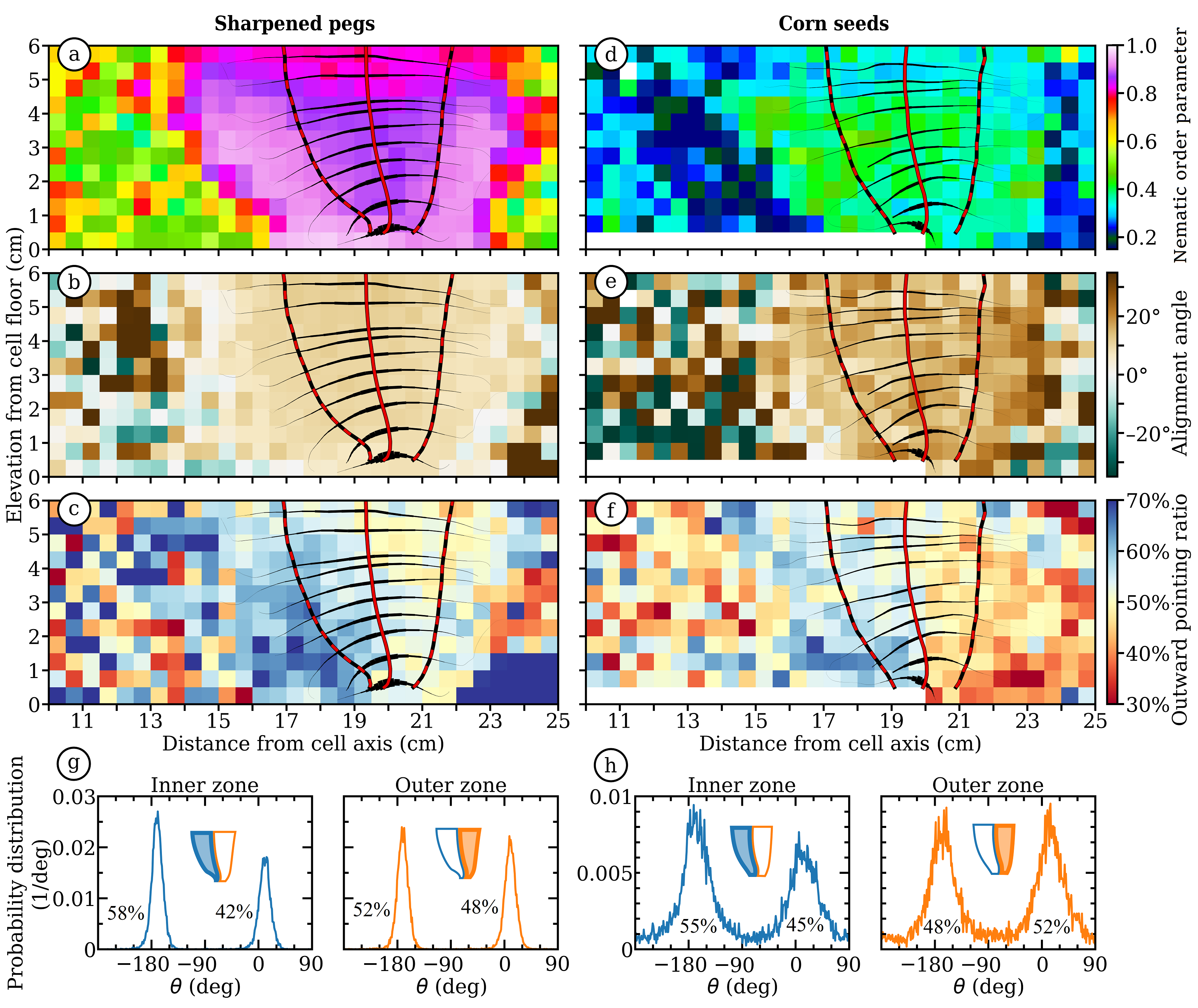}
\end{center}
  \caption{Maps of the orientational order parameter (row 1), average alignment angle (row 2)
ratio of outward pointing particles (row 3) and the distributions of particle orientations
(row 4) for sharpened pegs (left) and corn seeds (right) in the cylindrical split-bottom shear cell.
The contours and the middle of the shear zone core are indicated with red lines, the direction and strength of the velocity gradient is indicated with black lines.
  }
  \label{shear-rod-corn}
\end{figure}

As a first step, we consider the particles as simple elongated grains (i.e. in the first 
approximation we prescind from the difference between the non-sharpened and sharpened ends) 
and characterize the shear-induced alignment by a nematic order parameter $S$. This is calculated by diagonalizing the symmetric traceless order tensor \T{}:
\begin{equation}
T_{ij}= \frac{3}{2N} \sum\limits_{n=1}^N \left[{\ell}^{(n)}_i
{\ell}^{(n)}_j
-\frac{1}{3} \delta_{ij}
\right] \quad ,
\end{equation}
where $\vec {\ell}^{(n)}$ is a unit vector along the long axis of particle
$n$, and the sum is over all $N$ detected particles in a certain volume element within the container.
The largest eigenvalue of \T{} is the primary nematic order parameter $S$ \cite{borzsonyiPRL2012,borzsonyiPRE2012}.
Random grain orientations would lead to $S=0$, while $S=1$ corresponds to a perfect alignment 
with all grains parallel to each other. 
Figs. \ref{shear-rod-corn}(a) and (d) show the maps of the nematic order parameter $S$, while
Figs. \ref{shear-rod-corn}(b) and (e) show the maps of the average alignment angle
for sharpened pegs and corn seeds, respectively, in dependence of the height and the distance from the central axis. The data are averaged over rings around the center, where the local coordinate system always has one coordinate in vertical direction and a second one directed radially. 
These figures clearly show that both systems get aligned in the sheared region, with the 
strongest alignment in the core of the shear zone. The order parameter reaches a larger value
for sharpened pegs ($S\approx0.9$) than for the less elongated corn seeds ($S\approx0.5$). 
The alignment angle is slightly smaller for sharpened pegs ($\theta_\text{av}\approx10^\circ$) 
than for corn seeds ($\theta_\text{av}\approx14^\circ$). All this is in accordance with earlier 
observations on normal (not sharpened) pegs and and rice particles \cite{borzsonyiPRL2012,borzsonyiPRE2012}.

As a second step, we determine the orientation distribution of the particles by fully considering 
their asymmetric shapes (i.e. taking into account the difference between their two ends).  
Figs. \ref{shear-rod-corn}(g,h) show the distribution of the orientation angle $\theta$,
defined as the angle between the long axis of the particle and the flow direction as shown in 
Fig. \ref{orientations}(b) (top view).
The orientation distribution is separately shown in the outer and inner half of the core of the 
shear zone for both grain types: sharpened pegs and corn.
The distributions have narrower peaks for pegs, i.e. shear induced ordering is stronger for 
more elongated grains, and the peaks are slightly shifted with respect to the flow direction 
($0^\circ$ and $180^\circ$), in accordance with earlier observations on other elongated particles 
\cite{borzsonyiPRL2012,borzsonyiPRE2012}.
Interestingly, in the outer half of the shear zone core the peaks are symmetric, while in the inner 
half of the shear zone core they are asymmetric. 
One method to quantify the asymmetry is to calculate the 
fraction of grains corresponding to each peak. This turns to be $58\%:42\%$ and $55\%:45\%$ 
in the inner half of the zone for sharpened pegs and corn, respectively 
(see Figs. \ref{shear-rod-corn}(g,h)). In the outer half of the zone we find $52\%:48\%$, 
which is consistent with a symmetric distribution within our statistical uncertainty.
An other method to quantify the asymmetry is to introduce the polar order parameter
$P = \abs{\langle\cos(\theta-\theta_\text{av})\rangle}$. This yields $P=0.16$ and $0.09$ in the inner half and $P=0.038$ and $0.023$ in the outer half of the shear zone core for sharpened pegs and corn, respectively.
In order to understand this asymmetry, we should consider that the shear 
gradient is nearly horizontal in the outer half, while it is slightly tilted in the inner half, 
as discussed above. Thus the deviation of the particle's orientation from the flow direction has a vertical component in the inner half, while it is nearly horizontal in the outer half.
The asymmetry of the distributions in the inner half means, that there is a higher 
probability to find grains with their sharper end pointing outwards than inwards. 
This can be 
explained by the argument, that the two configurations are energetically different, since outward 
pointing grains have their thicker end at a lower position, while inward pointing grains have 
their thicker end at a higher position (see the two grains presented as examples in 
Fig. \ref{orientations}(b) on the left side of the side view). For a horizontal shear gradient 
there is no such difference (see the two grains on the right side of the side view), so the 
distributions in the outer half of the zone (see again Figs. \ref{shear-rod-corn}(g,h)) 
are nearly symmetric.

The maps of the outward pointing ratio thus nicely visualize this asymmetry throughout the whole 
cell (see Figs. \ref{shear-rod-corn}(c,f)). The asymmetry is more pronounced towards the inner 
side of the shear zone (blue colors) and is larger for the case of sharpened pegs than for corn seeds.

\subsection{Clogging in a silo}

\begin{figure}[!ht]
\begin{center}
\includegraphics[width=0.8\columnwidth]{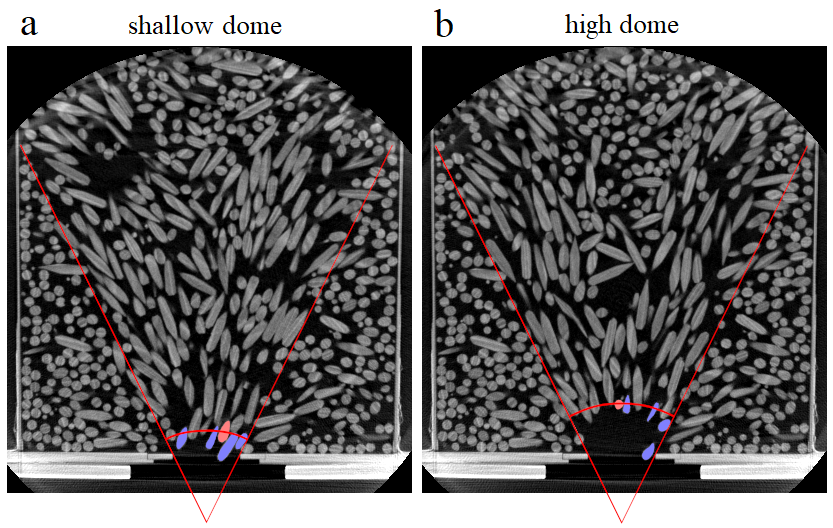}
\end{center}
\caption{
Vertical cross sections of two clogged configurations of sharpened pegs, and illustration of the 
definition of the dome. Straight red lines indicate the cone of moving particles. Within this cone, the
first 4 particles closest to the cone apex are marked blue; the 5$^{\rm th}$ closest, which is sitting on the dome surface (red arc),  is labeled light red. In the actual analysis, the 5$^{\rm th}$-closest particle within the complete 3D cone is selected, not just in a 2D cross section as in this illustration.
}
\label{clogged-pegs}
\end{figure}

In the second experiment, we investigate the grain orientations during the discharge of a nearly 
cylindrical silo with a circular orifice at the bottom. The size of the orifice was relatively small:
$D=3.3$ cm for sharpened pegs, and $D=2.6$, $2.8$ and $3.0$ cm for corn seeds.
This ensured frequent clogs during the discharge process. A typical avalanche size between two 
subsequent clogs corresponds to a handful of grains. A tomogram was recorded of each clogged 
configuration. 
The vertical cross sections of two example tomograms are shown in Fig. \ref{clogged-pegs}. 
As we see the clogs are not uniform, the figure shows two cases with a typical high and shallow dome.
During the avalanches the grains sink towards the orifice in a cone shaped region (indicated 
with red lines in Fig.~\ref{clogged-pegs}), while other grains in a stagnant zone near the 
silo wall do not move. The region with the strongest shear is nicely visualized by the shear induced orientational ordering of the grains, see the region with pink color in Fig. \ref{setup}(b), and it is also visible by the grain orientations in Fig.~\ref{clogged-pegs}. Grains in the stagnant zone (outside of the shear zone) are mostly horizontal, as they keep the orientation that they obtained during the filling procedure, similarly to earlier observations \cite{CruzHidalgo2010a}.
Our goal is to determine the particle orientations in the sheared region, and compare the statistics in the layer forming the dome and above. For this we need to define the approximate position of the dome for each tomogram. This is done by drawing a spherical cap centered around the cone apex, and going through the center of mass of the 5$^\text{th}$ closest particle to the apex. In Fig. \ref{clogged-pegs}
particles No. 1-4 are colored blue, while particle No 5. is colored red. Using the 5$^\text{th}$ closest particle instead of the closest one reduces the fluctuations caused by the position of a single particle.

\begin{figure}[!htp]
\begin{center}
\includegraphics[width=\columnwidth]{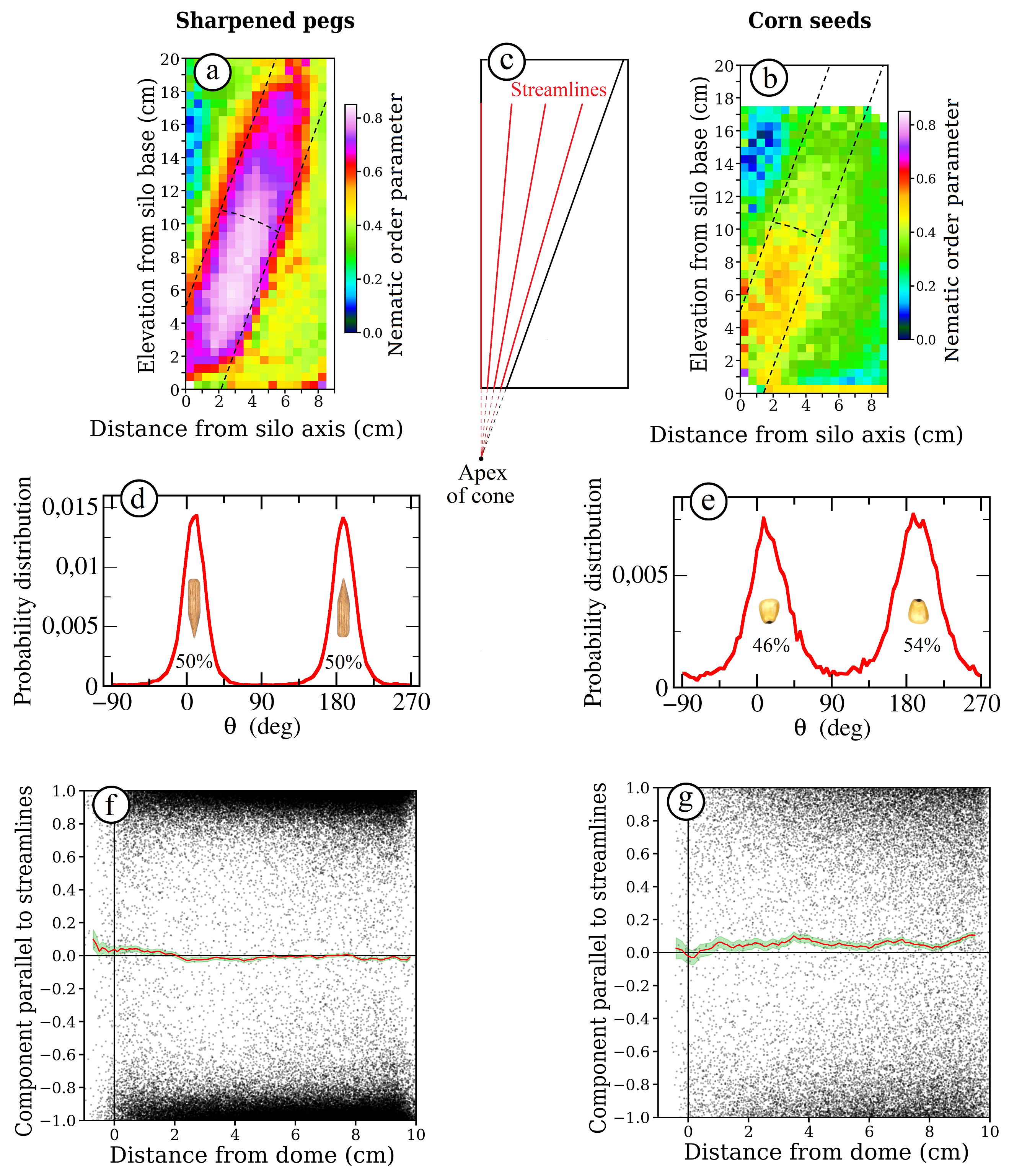}
\end{center}
\caption{
Grain orientation data for sharpened pegs and corn seeds in the silo.
Panels (a) and (b) show the order parameter maps, panel (c) shows the approximate  streamlines. 
Half of the silo cross section is shown, with the center at the left edge of the image and the 
outer wall at the right edge. Panels (d) and (e) show the orientation distribution of the 
particles' long axis with respect to the streamlines, while panels (f) and (g) quantify the 
orientations by showing the value of the component parallel to the streamlines as a function 
of the particles distance from the dome which forms the clog above the silo outlet. 
The red line corresponds to the average of the data, while the green band indicates its uncertainty. 
}
\label{orientations-in-silo-rod-corn}
\end{figure}

In order to identify the sheared region of our granular sample in the silo, we plot the
map of the orientational order parameter in a vertical cross section of the silo 
(see Figs. \ref{orientations-in-silo-rod-corn}(a,b)) by averaging the data in the azimuthal 
direction. As we see, orientational ordering is observed in a tilted region with a much larger
order parameter for sharpened pegs than for corn seeds. This is similar to the observations 
in the split bottom shear cell described above. In the following, we analyze particle orientations 
in the regions in between the dashed lines in Figs. \ref{orientations-in-silo-rod-corn}(a,b), 
up to about 10 cm above the orifice.
First we plot the distributions of the orientation angle $\theta$ measured with respect to the 
streamlines (see Figs. \ref{orientations-in-silo-rod-corn}(d,e)). Here the streamlines are 
approximated by straight lines converging towards the apex of the cone corresponding to 
the flowing region as shown in Fig. \ref{orientations-in-silo-rod-corn}(c).
As we see in Figs.~\ref{orientations-in-silo-rod-corn}(d,e), the grains are oriented nearly 
parallel to the flow with a slight majority of tip up particles ($54\%$) for corn and a practically equal fraction of tip up/down particles for sharpened pegs. 
The related polar order parameters are 0.06 and 0.003, respectively.
For the case of corn, the asymmetry 
in the silo is similar to the asymmetry observed in the inner zone of the shear cell 
($55\%:45\%$, see Fig. \ref{shear-rod-corn}(h)).
But for the case of sharpened pegs, the global symmetric distribution observed in the silo
is very different from the asymmetric one ($58\%:42\%$, Fig. \ref{shear-rod-corn}(g)) 
observed in the shear cell. 
This difference can be related to the fact, that the time period of rotation of a more elongated
particle in a shear flow is larger. In the shear cell we ensured a stationary state with extensive 
shearing of the sample, but in the silo the sample is subjected to a limited shear deformation, 
which is not enough to reach a stationary state for the relatively long pegs. 

We continue the analysis by looking for differences between the orifice region 
(where the clog occurs) and above. Namely, it might be expected that clogs are formed 
with a higher probability with "wedge down" particles. 
For this analysis we chose a different 
visualization of the data characterizing the orientation of the long axis of the particles: 
in Figs. \ref{orientations-in-silo-rod-corn}(f,g), we show the value of the component 
parallel to the streamlines as a function of the particles distance from the dome, which 
forms the clog above the silo outlet. The component parallel to the streamlines is 
positive when the sharper end of the particle points away from the orifice and it is negative when it points towards the orifice. In Figs. \ref{orientations-in-silo-rod-corn}(f) and (g) we see two 
clouds of points, one on the positive and one on the negative side of the graph. These clouds 
indicate the orientational ordering of the grains, and show stronger ordering for sharpened 
pegs than for corn seeds. The balance between the two clouds is visualized by the red data curve.
Red data points above zero mean more particles pointing up (i.e. away from the orifice) 
than down. 

As we see, for corn seeds the balance is slightly positive above the dome, but is 
around zero near the dome. This means that above the dome the wedge shaped corn seeds slightly
favour the ``sharp end up'' configuration against the ``sharp end down'' configuration 
(in accordance with the energetic argument described above), while in the dome, where the clog 
is formed, this difference seems to disappear. This is coherent with the argument that a dome 
forms more easily with wedge shaped particles pointing down.
Note that the distributions shown represent the clogged state. When the system is in free flow, 
the distributions of the polar axes in the orifice region may differ.

The scenario is different for the case of sharpened pegs. There, the balance is practically 
zero in the upper region, meaning equal number of ``sharp end up'' and ``sharp end down'' particles.
In the lower part (including the dome region) however, the balance becomes slightly positive, 
meaning slightly more ``sharp end up'' particles. Such distribution of particle orientations is 
coherent with our previous observations in the same silo on similar pegs without sharpened 
ends \cite{borzsonyiNJP2016}. 
Namely, the pegs with elongation $L/d=5$ gradually developed orientational ordering during 
sinking in the silo, and the orientation distribution did not fully reach the distribution 
observed in stationary shear flow even in their lowest position near the silo exit. 
The red curve in Fig. \ref{orientations-in-silo-rod-corn}(f) thus indicates two things: 
(i) due to the limited amount of shear in the course of silo flow, the sharpened 
pegs only start exploring the energetically more favoured ``sharp end up'' configuration in 
the lower part of the silo, 
(ii) since their fat side has a cylindrical shape (i.e it is not a wedge) the tendency to form a 
clogging arch rather with a ``sharp end down'' is not observed.
Altogether, although the difference between the case of sharpened pegs and corn seeds is small, 
it indicates that the particle configurations and  the mechanism for clog formation is different 
for these two types of particles.

\section{Conclusion}\label{conclusion}

Our experimental observations on the flow of asymmetric elongated particles  in a sheared system 
and during silo discharge show, that for both corn grains and pegs sharpened at one end
one can detect the effect of the particle asymmetry on the shear induced orientational ordering 
process. Namely, elongated grains in general rotate in a shear flow with modulated rotation velocity:
faster rotation when perpendicular to the flow and slowing down when their orientation is near 
the flow direction. This leads to orientational ordering, as grains spend more time with their 
long axis oriented near the flow direction. This general behavior of elongated grains is slightly 
modified for the asymmetric particles investigated here: if their plane of rotation is not 
perpendicular to gravity, they are expected to spend 
more time in the ``sharper end up'' than in the ``sharper end down'' configuration, so the 
orientational distribution is expected to become asymmetric. We experimentally detect this 
asymmetry in the particle orientations in a cylindrical split bottom shear cell for both corn 
seeds and sharpened pegs. For the case of silo discharge, the asymmetry is detected in most of the 
sheared region for corn seeds, except in the vicinity of the orifice. This indicates, that for 
wedge shaped particles clog formation is slightly enhanced when more ``sharper end down'' 
particles are present in the region where the clogging arch is formed. For sharpened pegs the
asymmetry only appears in the orifice region with a small amplitude, indicating that (i) for longer 
grains, larger shear deformation is needed to reach the stationary state and (ii) for sharpened 
pegs (which are not true wedges, but have a cylindrical part at the thick side) clog formation 
does not seem to be influenced by the particles asymmetry. Future numerical simulations might 
help to get more insight into the processes leading to the asymmetric orientational distributions 
observed in our experiments.

\section*{Acknowledgements}
Financial support by the European Union's Horizon 2020 Marie Sk\l{}odowska-Curie grant ''CALIPER'' 
(No. 812638), by the Hungarian National Research, Development and Innovation Office -- 
NKFIH (Grant No. OTKA K 116036), by the DAAD/TKA researcher exchange program (Grant No. 274464) 
are acknowledged. The authors thank G. Rose for providing access to the X-ray CT facility and 
B. Szab\'o \ for technical assistance. B. F. and T.B appreciate discussions with  J. Dijksman 
and J. van der Gucht.

\section*{References}

\def\newblock{\ }
\bibliographystyle{unsrt}

\end{document}